%% file: main.tex
\documentclass[twocolumn]{aastex631}

\usepackage{graphicx}	
\usepackage{amsmath}	
\usepackage{amssymb}	
\usepackage{xcolor}
\usepackage{subfigure}

\include{defns}

\newcommand{\xmm}{{ XMM-Newton }}

\newcommand{\Rg}{\ensuremath{R_\mathrm{g}}}

\begin{document}
\title[Mrk~335 Reverberation Mapping]{UV/Optical disk reverberation lags despite a faint X-ray corona in the AGN Mrk~335}
\correspondingauthor{Erin Kara}
\email{ekara@mit.edu}


\author[0000-0003-0172-0854]{Erin Kara}
\affiliation{MIT Kavli Institute for Astrophysics and Space Research, Massachusetts Institute of Technology, Cambridge, MA 02139, USA}

\author[0000-0002-3026-0562]{Aaron J.\ Barth}
\affiliation{Department of Physics and Astronomy, 4129 Frederick Reines Hall, University of California, Irvine, CA, 92697-4575, USA}

\author[0000-0002-8294-9281]{Edward M.\ Cackett}
\affiliation{Department of Physics and Astronomy, Wayne State University, 666 W.\ Hancock St, Detroit, MI, 48201, USA}

\author[0000-0001-9092-8619]{Jonathan Gelbord}
\affiliation{Spectral Sciences Inc., 4 Fourth Ave., Burlington, MA 01803, USA}

\author[0000-0001-5639-5484]{John Montano}
\affiliation{Department of Physics and Astronomy, 4129 Frederick Reines Hall, University of California, Irvine, CA, 92697-4575, USA}

\author[0000-0001-5841-9179]{Yan-Rong Li}
\affiliation{Key Laboratory for Particle Astrophysics, Institute of High Energy Physics, Chinese Academy of Sciences, 19B Yuquan Road,\\ Beijing 100049, People's Republic of China}

\author[0000-0003-3907-8000]{Lisabeth Santana}
\affiliation{Department of Physics and Astronomy, University of Pittsburgh, 3941 O'Hara St, Pittsburgh, PA 15260, USA}

\author[0000-0003-1728-0304]{Keith Horne}
\affiliation{SUPA School of Physics and Astronomy, North Haugh, St.~Andrews, KY16~9SS, Scotland, UK}



\author[0000-0003-2658-6559]{William N. Alston}
\affiliation{European Space Agency (ESA), European Space Astronomy Centre (ESAC), Villanueva de la Canada, Madrid, E-28691, Spain}

\author{Douglas Buisson}
\affiliation{Independent}

\author[0000-0002-4830-7787]{Doron Chelouche}
\affiliation{Department of Physics, Faculty of Natural Sciences, University of Haifa, Haifa 3498838, Israel}
\affiliation{Haifa Research Center for Theoretical Physics and Astrophysics, University of Haifa, Haifa 3498838, Israel }

\author{Pu Du}
\affiliation{Key Laboratory for Particle Astrophysics, Institute of High Energy Physics, Chinese Academy of Sciences, 19B Yuquan Road,\\ Beijing 100049, People's Republic of China}

\author[0000-0002-9378-4072]{Andrew C. Fabian}
\affiliation{Institute of Astronomy, 
University of Cambridge, 
Madingley Road, Cambridge CB3 0HA, UK}

\author{Carina Fian}
\affiliation{School of Physics and Astronomy and Wise observatory, Tel Aviv University, Tel Aviv 6997801, Israel}
\affiliation{Haifa Research Center for Theoretical Physics and Astrophysics, University of Haifa, Haifa 3498838, Israel }

\author{Luigi Gallo}
\affiliation{Department of Astronomy and Physics, Saint Mary’s University, 923 Robie Street, Halifax, NS, B3H 3C3, Canada}

\author[0000-0002-2908-7360]{Michael R.\ Goad}
\affiliation{School of Physics and Astronomy, University of Leicester, University Road, Leicester, LE1 7RH, UK}

\author{Dirk Grupe}
\affiliation{Department of Physics, Geology, and Engineering Technology, Northern Kentucky University, 1 Nunn Dr., Highland Heights, KY 41099}

\author[0000-0002-9280-1184]{Diego H.\ Gonz\'{a}lez Buitrago}
\affiliation{Instituto de Astronom\'{\i}a, Universidad Nacional Aut\'{o}noma de M\'{e}xico, Km 103 Carretera Tijuana-Ensenada, 22860 Ensenada B.C., M\'{e}xico}

\author[0000-0002-6733-5556]{Juan V.\ Hern\'{a}ndez Santisteban}
\affiliation{SUPA School of Physics and Astronomy, North Haugh, St.~Andrews, KY16~9SS, Scotland, UK}

\author[0000-0002-9925-534X]{Shai Kaspi}
\affiliation{School of Physics and Astronomy and Wise observatory, Tel Aviv University, Tel Aviv 6997801, Israel}

\author{Chen Hu}
\affiliation{Key Laboratory for Particle Astrophysics, Institute of High Energy Physics, Chinese Academy of Sciences, 19B Yuquan Road,\\ Beijing 100049, People's Republic of China}

\author{S. Komossa} 
\affiliation{Max-Planck-Institut für Radioastronomie, Auf dem Hügel 69, 53121 Bonn, Germany}

\author[0000-0002-2180-8266]{Gerard A.\ Kriss}
\affiliation{Space Telescope Science Institute, 3700 San Martin Drive, Baltimore, MD 21218, USA}

\author[0000-0002-8671-1190]{Collin Lewin}
\affiliation{MIT Kavli Institute for Astrophysics and Space Research, Massachusetts Institute of Technology, Cambridge, MA 02139, USA}

\author[0000-0002-9854-1432]{Tiffany Lewis}
\affiliation{NASA Postdoctoral Program Fellow, Astroparticle Physics Lab, Goddard Space Flight Center}
\affiliation{Astrophysics Science Division, 
NASA Goddard Space Flight Center, 8800 Greenbelt Road, 
Greenbelt, MD 20771, USA}

\author[0000-0002-1661-4029]{Michael Loewenstein}
\affiliation{Astrophysics Science Division, 
NASA Goddard Space Flight Center, 8800 Greenbelt Road, 
Greenbelt, MD 20771, USA}
\affiliation{Department of Astronomy, 
University of Maryland, 
College Park, MD 20742, USA}

\author{Anne Lohfink}
\affiliation{Department of Physics, Montana State University, P.O. Box 173840, Bozeman, MT 59717-3840, USA}

\author[0000-0003-4127-0739]{Megan Masterson}
\affiliation{MIT Kavli Institute for Astrophysics and Space Research, Massachusetts Institute of Technology, Cambridge, MA 02139, USA}

\author{Ian M.\ M$^{\rm c}$Hardy}
\affiliation{School of Physics and Astronomy, University of Southampton, Highfield, Southampton SO17 1BJ, UK}

\author[0000-0002-4992-4664]{Missagh Mehdipour}
\affiliation{Space Telescope Science Institute, 3700 San Martin Drive, Baltimore, MD 21218, USA}

\author{Jake Miller}
\affiliation{Department of Physics and Astronomy, Wayne State University, 666 W.\ Hancock St, Detroit, MI, 48201, USA}

\author{Christos Panagiotou}
\affiliation{MIT Kavli Institute for Astrophysics and Space Research, Massachusetts Institute of Technology, Cambridge, MA 02139, USA}

\author{Michael L.\ Parker}
\affiliation{Institute of Astronomy, 
University of Cambridge, 
Madingley Road, Cambridge CB3 0HA, UK}

\author[0000-0003-2532-7379]{Ciro Pinto}
\affiliation{INAF - IASF Palermo, 
Via U. La Malfa 153, 
I-90146 Palermo, Italy}

\author{Ron Remillard}
\affiliation{MIT Kavli Institute for Astrophysics and Space Research, Massachusetts Institute of Technology, Cambridge, MA 02139, USA}

\author{Christopher Reynolds}
\affiliation{Institute of Astronomy, 
University of Cambridge, 
Madingley Road, Cambridge CB3 0HA, UK}

\author{Daniele Rogantini}
\affiliation{MIT Kavli Institute for Astrophysics and Space Research, Massachusetts Institute of Technology, Cambridge, MA 02139, USA}

\author[0000-0001-9449-9268]{Jian-Min Wang}
\affiliation{Key Laboratory for Particle Astrophysics, Institute of High Energy Physics, Chinese Academy of Sciences, 19B Yuquan Road,\\ Beijing 100049, People's Republic of China}
\affiliation{School of Astronomy and Space Sciences, University of Chinese Academy of Sciences, 19A Yuquan Road, Beijing 100049, People's Republic of China}
\affiliation{National Astronomical Observatories of China, 20A Datun Road, Beijing 100020, People's Republic of China}

\author[0000-0002-1742-2125]{Jingyi Wang}
\affiliation{MIT Kavli Institute for Astrophysics and Space Research, Massachusetts Institute of Technology, Cambridge, MA 02139, USA}

\author[0000-0002-4794-5998]{Dan Wilkins}
\affiliation{Kavli Institute for Particle Astrophysics and Cosmology, Stanford University, 452 Lomita Mall, Stanford, CA 94305, USA}

\begin{abstract}

We present the first results from a 100-day Swift, NICER and ground-based X-ray/UV/optical reverberation mapping campaign of the Narrow-Line Seyfert 1 Mrk~335, when it was in an unprecedented low X-ray flux state. Despite dramatic suppression of the X-ray variability, we still observe UV/optical lags as expected from disk reverberation. Moreover, the UV/optical lags are consistent with archival observations when the X-ray luminosity was $>10$ times higher. Interestingly, both low- and high-flux states reveal UV/optical lags that are $6-11$ times longer than expected from a thin disk. These long lags are often interpreted as due to contamination from the broad line region, however the $u$ band excess lag (containing the Balmer jump from the diffuse continuum) is less prevalent than in other AGN. The Swift campaign showed a low X-ray-to-optical correlation (similar to previous campaigns), but NICER and ground-based monitoring continued for another two weeks, during which the optical rose to the highest level of the campaign, followed $\sim10$ days later by a sharp rise in X-rays. While the low X-ray countrate and relatively large systematic uncertainties in the NICER background make this measurement challenging, if the optical does lead X-rays in this flare, this indicates a departure from the zeroth-order reprocessing picture.  If the optical flare is due to an increase in mass accretion rate, this occurs on much shorter than the viscous timescale. Alternatively, the optical could be responding to an intrinsic rise in X-rays that is initially hidden from our line-of-sight.

\end{abstract}


\section{Introduction}

The accretion of gas onto supermassive black holes is a key driver in understanding the formation and evolution of galaxies. Most of the accretion power is released very close to the black hole, in the form of radiation and large-scale outflows. Understanding the inner accretion flow--at the intersection of inflow and outflow--is essential for understanding the Active Galactic Nucleus (AGN) feedback phenomenon. While these scales are generally too small to be spatially resolved, reverberation light echoes offer us a way to infer the geometry and dynamics of gas flows close to the black hole.

As gas funnels in towards the supermassive black hole (SMBH) at the center of a galaxy, collisions and angular momentum conservation cause the formation of an optically thick, geometrically thin accretion disk around the black hole \citep{SS1973} that emits thermal radiation in the optical/near infrared (NIR) and peaking in the ultraviolet (UV) band. Some of those UV photons scatter off mildly relativistic electrons in a region close to the black hole that is known as the corona \citep{haardt91}. This scattering boosts the UV photons to X-ray energies. A fraction of the X-ray photons emitted by the corona reach the observer directly, while other photons first intercept the accretion disk, are reprocessed, and then re-emitted. The energetic photons interact with the material in the disk via processes such as photoelectric absorption and fluorescence, thermalization, and Compton scattering. The resulting reprocessed emission that is produced in the X-ray band is often referred to as  the `reflection spectrum',  \citep[e.g.,][]{fabian89,ross05,garcia10}, but much emission is reprocessed into the UV-optical-infrared (UVOIR) band, as well, due to disk heating.

Both the X-ray reflection spectrum and the broadband UVOIR SED from reprocessed disk emission can be used to infer the geometry and composition of the accretion disk, from scales of $10^{2-4}$\Rg{}\footnote{ \Rg{}$=GM/c^{2}$}, as probed in the UVOIR, and down to $1-10$\Rg{}, as probed in X-rays. X-ray spectra can be particularly powerful because atomic features in the reflection spectrum are observed to be broadened due to Doppler motion in the accretion disk, and also by the gravitational redshift from the strong potential well of the black hole.

Complementary to spectral information, the time dependence of photons originating in the innermost regions of the AGN can map out the accretion disk, and very importantly, provide insights into the {\em physical} size scales around the black hole (e.g. in centimeters, rather than in units of $GM/c^{2}$). In the past decade, advances in higher cadence observing campaigns and improved analysis techniques have allowed us to measure the light travel time delays between the continuum emission from the innermost regions and the reprocessed emission off the disk that shines in X-ray and UVOIR. Typical time delays in X-ray reverberation lags are tens to hundreds of seconds (corresponding to light travel distances of tens of gravitational radii), and in UVOIR continuum reverberation mapping time delays are on the order of days (probing the accretion disk at larger distances of hundreds to thousands of gravitational radii). See \citet{cackett21} for a recent review of AGN reverberation mapping on all scales from the dusty torus ($>10^{5}$\Rg{}) to the broad line region (BLR; $10^{3-5}$\Rg{}) and down to the innermost stable circular orbit (ISCO; $\sim 1$\Rg{}).


\begin{figure*}
    \centering
    \includegraphics[width=\textwidth]{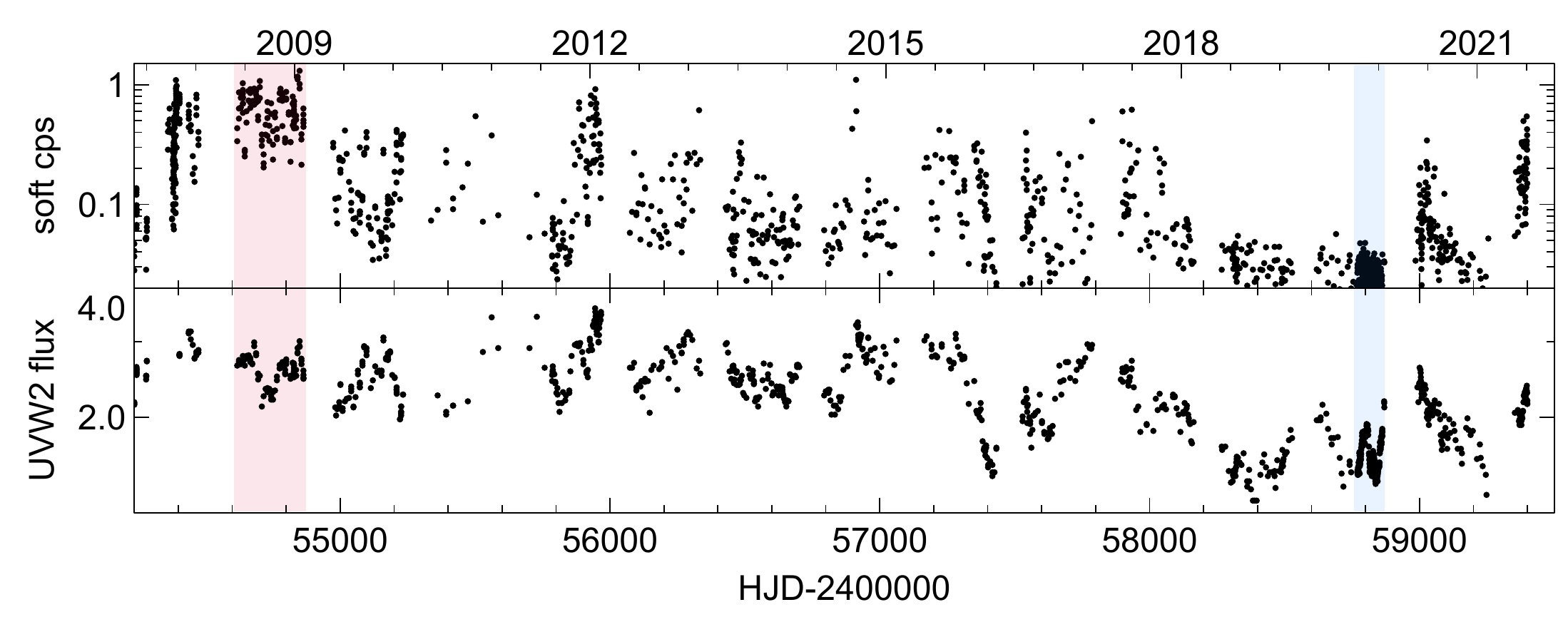}
    \caption{The long-term Swift light curve from 2007 to 2021, showing the 0.3--1.5~keV light curve (top), and the UVW2 light curve (bottom). Our high-cadence disk reverberation mapping campaign in 2019--2020 took place in an unprecedentedly low X-ray flux state (blue). Later we compare the lags to a lower-cadence campaign taken during a high flux epoch (red). The soft band light curve is in units of counts~s$^{-1}$ (note log scale), and the UVW2 light curve is in units of $10^{-14}$~erg~cm$^{-2}$~s$^{-1}$\AA$^{-1}$.}
    \label{fig:swiftlc}
\end{figure*}

High-cadence X-ray/UV/optical continuum reverberation mapping campaigns have successfully been successfully completed for nine AGN (NGC 5548: \citealt{Edelson2015,fausnaugh16}, NGC 4151: \citealt{edelson17}, NGC 4593: \citealt{McHardy18,cackett18}, Mrk 509: \citealt{edelson19}, Fairall 9 \citealt{Hernandez2020}, Mrk 142: \citealt{cackett20}, Mrk 110: \citealt{vincentelli21}, Mrk 817: \citealt{kara2021}, Mrk 335: this work), and several more campaigns have recently been completed or are currently underway. While the sample is yet small, the campaigns have revealed surprising results that present challenges to standard accretion theory. In nearly all AGN that have been the subject of these high-cadence multi-wavelength campaigns, the amplitude of the X-ray-to-UV lag does not match the standard model (see, e.g., \citealt{edelson19}). Moreover, the correlation between X-ray and UV is much weaker than predicted from the standard reprocessing model. Some have suggested that this means that the observed X-rays are not the driving light curve at all, or that the thin accretion disk does not extend to the ISCO, but instead, truncates at some radius ($\sim 20$\Rg{}), within which, there is a hot inner flow that emits a soft X-ray continuum \citep{Gardner2017,Mahmoud20}.  This is in tension with results from X-ray reverberation, indicating that the disk is not highly truncated \citep{demarco13,kara16}. Yet others have suggested that by filtering out long timescale variability (that could be due to obscuration or the inflow of matter through the accretion disk), one can recover the expected X-ray to UV reverberation time lags \citep{McHardy18,Hernandez2020}.

In this paper we examine the X-ray/UVOIR disk reverberation lags in the well-known and rapidly variable narrow-line Seyfert 1 galaxy Mrk~335 ($z=0.025785$; \citealt{redshift}), which has been the subject of several X-ray, UV and optical timing campaigns. In a 120-day ground-based photometric and spectroscopic campaign, \citet{grier12} found the optical continuum to lead the H$\beta$ broad emission line by $13.9 \pm 0.9$ days, inferring a virial mass of $(2.6 \pm 0.8) \times 10^{7} M_{\odot}$. In an archival analysis of a long XMM-Newton observation, \citet{Kara2013} found that the X-ray continuum led the iron~K and soft excess reflection features by $\sim 100$~s. Recently, \citet{mastroserio20} modelled these X-ray lags with general relativistic ray-tracing simulations. Assuming a geometry of a point-source corona close to the black hole, they inferred a black hole mass that was ten times smaller than that found in the optical. These authors suggest that this inconsistency could be due to a more complex coronal geometry that is not accounted for in the modeling.

In addition to its rapid variability, Mrk~335 has shown dramatic, long-term changes, revealed by long-term Swift X-ray/UV monitoring (\citealt{grupe07, grupe12, gallo18, komossa20} and Fig.~\ref{fig:swiftlc}). While there do appear to be some long-term trends between the UV/optical and X-rays (e.g. \citealt{gallo18,parker19}), generally, the amplitude of the X-ray variability is much more dramatic, and can dip into extreme low-flux states, while the optical/UV remain persistently variable. The origin of these dramatic X-ray dips is still not well understood, sometimes being attributed to a collapse of
the corona (e.g. \citealt{gallo13,parker14,wilkins15}) or to absorption and partial obscuration of the disk and corona (e.g. \citealt{grupe08, Longinotti13, komossa20}).

Here we report on the results of a 100-day high-cadence campaign of Mrk~335 during its most recent extreme X-ray low-flux state, using the Neil Gehrels Swift Observatory (Swift), the Neutron Star Interior Composition ExploreR (NICER) on the International Space Station, and several ground-based observatories. In Section~\ref{sec:obs}, we describe details of the campaign and data reduction. We present the results of the multi-wavelength cross-correlation analysis in Section~\ref{sec:results}, and compare these results to archival high-flux observations taken in 2008. Both high- and low-flux state exhibit UV/optical lags as expected from the standard X-ray reprocessing model, and we present possible explanations in the Discussion (Section~\ref{sec:discuss}). Conclusions are summarized in Section~\ref{sec:conclusions}.

\section{Observations and Data Reduction}
\label{sec:obs}

\subsection{Neil Gehrels Swift Observatory}

Our campaign of Mrk~335 began on 2019 October 14 for 100 consecutive days with a cadence of roughly 3 visits per day, until 2020 January 22 (HJD-2450000=8770--8870). There were occasional gaps due to poor visibility or interruptions caused by  Targets of Opportunity. In particular, there was a large gap at the end of the campaign (HJD-2450000=8863--8869), due to gravitational wave counterpart searches. Each visit is typically $\sim$1 ksec. The Swift XRT \citep{burrows05} was operated in Photon Counting mode. The UVOT \citep{roming05} was typically operated in an end-weighted filter mode (0x224c) to get exposures in all 6 UV/optical filters with an exposure weighting of 3:1:1:1:1:2 (for UVW2 through V). This has the effect of greatly improving the V band signal-to-noise at the cost of slight degradation in the UVM2 to B band signals. This is done because the UVW2/V lag covers the largest wavelength range, and thus is the most physically constraining.

Swift X-ray light curves were generated using the Swift-XRT \citep{evans07,evans09} data product tool\footnote{\url{https://www.swift.ac.uk/user_objects/index.php}}.
All archival and new UVOT data were processed and analyzed following the procedures described by \citet{edelson19} and \citet{Hernandez2020}, with HEASOFT version 6.28 and CALDB version 20210113.  Fluxes are measured using the uvotsource tool, with a circular source extraction region of 5\arcsec\ radius and with the background measured in a surrounding 40\arcsec--90\arcsec\ annulus. We apply detector masks to reject data points when the source falls on regions of the chip with lower sensitivity.  We follow the procedure laid out in \citet{Hernandez2020}, but find that the detector masks employed there are too aggressive for the present data, eliminating many points that are consistent with the light curves within their measurement errors.  Instead, we use a more conservative set of masks defined by applying higher thresholds to the sensitivity maps.

The Swift X-ray count rate during the campaign was very low compared to archival observations from 2008-2018.  The mean 0.3 -- 10 keV count rate between 2007 -- 2018 is 0.26~c\,s$^{-1}$, while during our 100-day campaign the mean rate is 0.038~c\,s$^{-1}$, a factor of 7 lower.  Despite the significant change in the X-ray count rate, the mean UVW2 flux dropped only by 12\% during our campaign (to $2\times10^{-14}$ erg~s$^{-1}$~cm$^{-2}$~\AA$^{-1}$). Because the Swift X-ray count rate is very low, we binned the X-ray light curves to a 3-day binning. This reduces the scatter, but the resultant X-ray to UVW2 lags are the same regardless of bin choice. Later, in Section~\ref{sec:results}, we compare the lags in our low-state campaign, to a relatively high-cadence high-state archival year-long campaign in 2009 (red shaded region in Fig.~\ref{fig:swiftlc}). In this archival epoch, the X-rays were 23 times higher than during our low-flux campaign, while the UVW2 was only a factor of 1.3 times higher.

\subsection{NICER}

\begin{figure}
    \centering
    \includegraphics[width=\columnwidth]{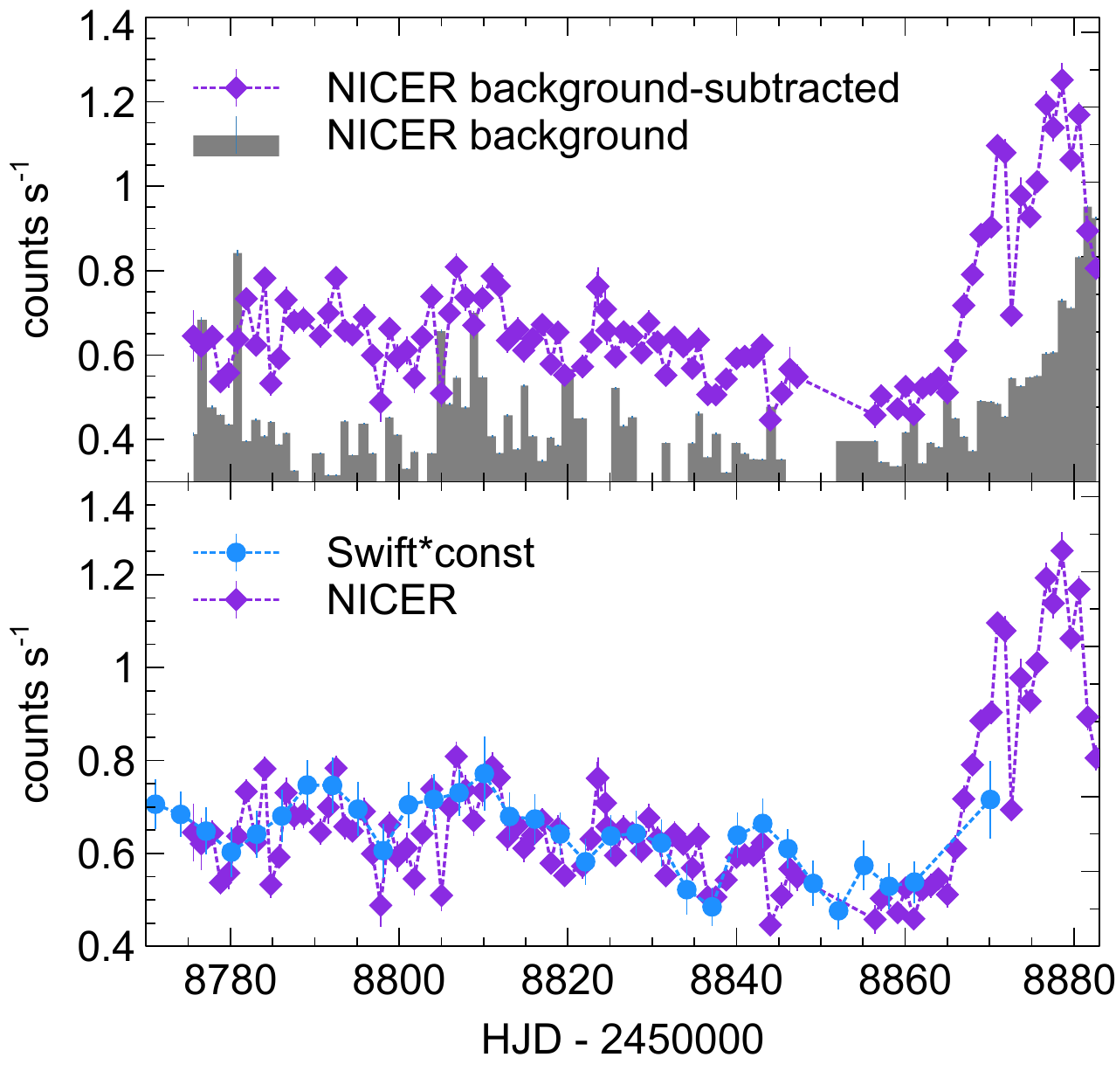}
    \caption{({\em Top:}) Background-subtracted 0.3--1.5~keV NICER light curve compared to the background rate (starting 2019 October 14). The increase in background at the end of the campaign is due to increased optical light leakage towards the end of the visibility window. Despite this, one can see an increase in the source light curve before the increase in background. The Swift light curve ({\em bottom}) also shows an increase in flux at the end of the campaign.}
    \label{fig:nicer_lc}
\end{figure}

\begin{deluxetable*}{llcl}
\label{table:optimg}
\tablecaption{Ground-based Imaging Observations}
\tablehead{
\colhead{Observatory/Telescope} & \colhead{Instrument} & \colhead{Filters}  & \colhead{References} }
\startdata
Las Cumbres Observatory 1m network   & Sinistro & $g^\prime r^\prime i^\prime z_\mathrm{s}$  & \citet{Brown2013} \\
Liverpool Telescope 2m & IO:O & $g^\prime r^\prime i^\prime z^\prime$ &  \citet{Steele2004} \\
San Pedro M\'{a}rtir Observatory 1.5m & RATIR &  $g^\prime r^\prime i^\prime$ &  \citet{Butler2012}, \citet{Watson2012} \\
Wise Observatory 18-inch & QSI683 &  $g^\prime r^\prime i^\prime z^\prime$ &  \citet{Brosch2008} \\
Zowada Observatory 20-inch & & $g^\prime r^\prime i^\prime z_\mathrm{s}$ &  \citet{Carr2022} \\
\enddata
\end{deluxetable*}

\begin{deluxetable}{lllll}
\label{table:photdata}
\tablecaption{Ground-Based Photometry}
\tablehead{\colhead{Filter} & \colhead{HJD-2450000} & \colhead{$f_\lambda$} &
        \colhead{$\sigma(f_\lambda)$} & \colhead{Telescope}}
        \startdata
$g$ & 8700.902 & 6.551 & 0.009 & LCO-V37 \\
$g$ & 8704.920 & 6.385 & 0.015 & LCO-V39 \\
$g$ & 8708.755 & 6.285 & 0.009 & LCO-V37 \\
$g$ & 8715.890 & 5.958 & 0.010 & LCO-V37 \\
$g$ & 8722.919 & 5.686 & 0.005 & LCO-V37 \\
$g$ & 8726.759 & 5.657 & 0.074 & LCO-W87 \\
$g$ & 8728.451 & 5.611 & 0.037 & Zowada \\
$g$ & 8730.816 & 5.647 & 0.005 & LCO-V37 \\
$g$ & 8733.901 & 5.805 & 0.006 & LCO-V37 \\
$g$ & 8738.712 & 6.010 & 0.047 & LCO-W86 \\
        \enddata
        \tablecomments{Flux densities ($f_\lambda$) and uncertainties $(\sigma)$ are given in units of $10^{-15}$ erg cm$^{-2}$ s$^{-1}$ \AA$^{-1}$. For Las Cumbres Observatory (LCO), the Minor Planet Center telescope codes are listed to denote individual telescopes in the 1 m network. This table is published in its entirety in the machine-readable format.
      A portion is shown here for guidance regarding its form and content.}
\end{deluxetable}

NICER was set to accompany the same 100~days of observations as Swift, with a one-day cadence. However, due to the historic first all-female space walk on the International Space Station on 2019 October 18, NICER had to stow, and could not begin the daily monitoring until 2019 October 20 (HJD-2450000=8776). This was, in some ways, fortuitous, as this initial delay led to the later extension of NICER's campaign until HJD~2458883, during which time the X-rays rose to their highest flux level of the entire campaign. 

The data were processed using the NICER data-analysis software version 2017-06-01\_V005, allowing for undershoot rates up to 400 (standard screening otherwise). The source and background spectra were extracted for each observation. The background spectra were constructed with the background estimator, 3C50 \citep{remillard22}. We removed data from noisy detectors 14 and 34, and for each GTI, we removed data from any detector (FPM) that deviates more than 3-$\sigma$ from the average in the 0--0.2~keV band, as a way to reduce the impact of optical light leak. The count rates were scaled to be the count rates per 52-FPMs, accounting for the detectors that were excluded or were off at a particular time. This results in 101 individual pointings. The source flux was so low that the background dominates above 1.5~keV, and therefore, we only examine the soft-band light curve with NICER. 

NICER's background is largely affected by optical light leakage, which increases depending on space weather and the proximity of bright objects (e.g., Sun, Earth, Moon, and their reflections on ISS solar panels). While we were fortunate that NICER was able to extend observations beyond the Swift campaign to when optical telescopes reached their peak luminosity, these NICER observations had a lower sun angle, and so this is also when NICER's background level rose to the highest of the campaign  (Fig.~\ref{fig:nicer_lc}). The inferred rise in source flux does not appear to be due to high background levels because (1) the X-ray source flux rises before the background does, (2) Swift also observed a rise in X-rays (albeit with very sparse sampling), and (3) the NICER spectrum during the flare (see Section~\ref{sec:spec}) clearly looks like the spectrum of Mrk~335 (i.e. with a soft continuum and emission lines from circumnuclear material), and not like the particle background spectrum, which is much harder ($\Gamma \sim 1$). See Section~\ref{sec:spec} for more details of the NICER spectrum during the flare state.

\subsection{XMM-Newton}

XMM-Newton observed Mrk~335 once, starting on HJD~2458845 for 105~ks during the 100-day reverberation mapping campaign through an XMM-Newton Director's Discretionary Time (DDT) observation (PI: N. Schartel; \citealt{tripathi20,liu2021}). 
We reduced the data using the XMM-Newton Science Analysis System (SAS v. 19.0.0) and the newest calibration files. We started with the observation data files and followed standard procedures. The PN observations were taken in Full Frame Mode. The source extraction regions are  circular regions of radius 35 arcsec centered on source position. The background regions are also circular regions of radius 35 arcsec or greater, avoiding other bright sources and the detector edges where the instrumental copper line is most prominent. The observations had some distinct soft proton background flares, which we removed by screening based on the 13--15~keV light curve count rate, resulting in a net exposure of 70~ks. The response matrices were produced using rmfgen and arfgen in SAS.

\begin{figure}
    \centering
    \includegraphics[width=\columnwidth]{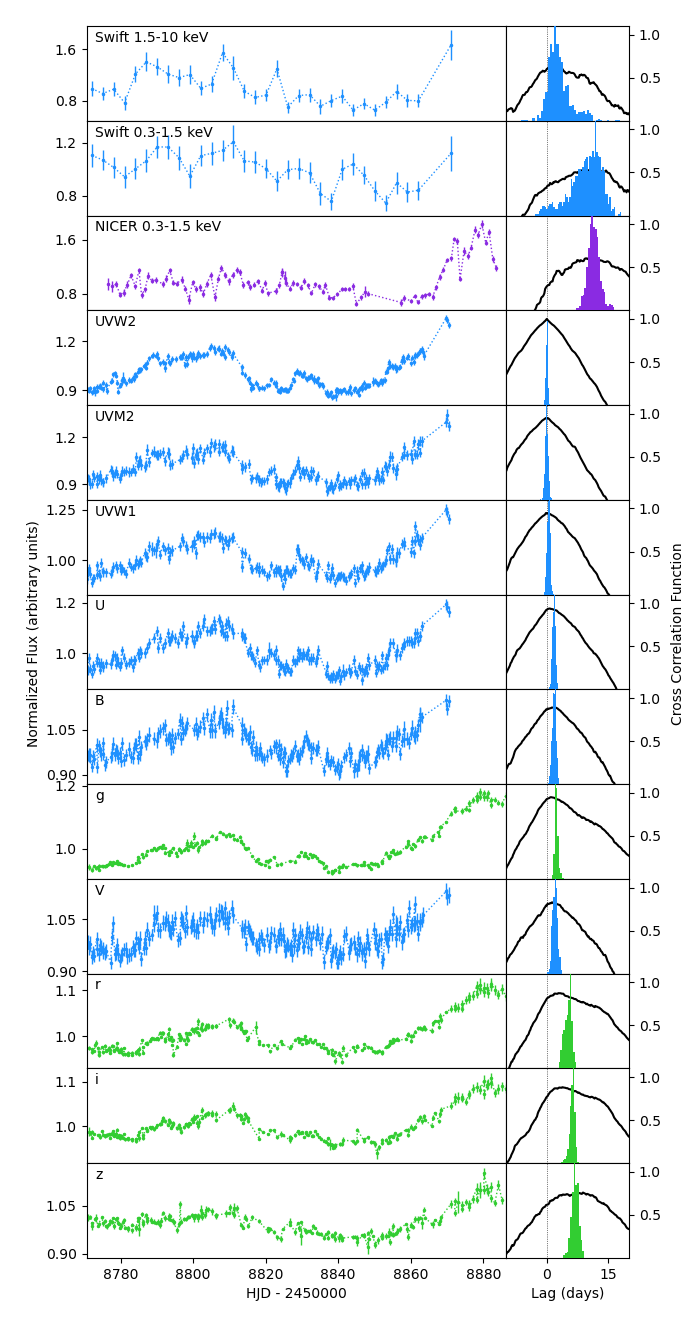}
    \caption{{\it Left:} Mrk~335 light curves from the high-cadence reverberation mapping campaign (starting 2019 October 14) with Swift (blue), NICER (purple), and ground-based telescopes (green). 
    The portions of the ground-based light curves before the start date of the Swift campaign are not shown. {\it Right:} Solid black lines show the cross-correlation function with respect to the Swift UVW2 band. A positive lag indicates the band of interest is lagging behind the UVW2 reference band. Colored histograms show the ICCF centroid lag distributions from the FR/RSS technique. See text for details.}
    \label{fig:contlc}
\end{figure}

\subsection{Ground-based photometry}

Ground-based optical imaging observations were obtained at several sites, listed in Table \ref{table:optimg}. The ground-based campaign spanned a longer duration than the Swift program, beginning on 2019 July 29 and continuing through 2020 February 13. Filters included the SDSS $g^\prime r^\prime i^\prime z^\prime$, and Pan-STARRS $z_s$. We combined the $z^\prime$ ($\lambda_c=9134$ \AA) and $z_s$ ($\lambda_c=8700$ \AA) bands for the final light curve measurements, and for simplicity we refer to the SDSS and Pan-STARRS filters as \emph{griz} bands hereinafter. We also obtained some data in the SDSS $u^\prime$ and Johnson \emph{BV} bands, but the temporal coverage of the \emph{BV} observations was poor compared with the SDSS bands, and the S/N of the $u^\prime$ data was too low to measure an accurate light curve. Since the Swift data includes the \emph{UBV} bands, we do not include the lower-quality ground-based $u^\prime$ and \emph{BV} data in this paper.
Initial data processing including bias and overscan subtraction and flat-fielding was carried out using the standard reduction pipeline for each telescope.

Aperture photometry was carried out using an automated pipeline based on AstroPy \citep{Astropy2018} routines, following the same methods used for a recent campaign on Mrk 817 as described by \citet{kara2021}. The photometric aperture radius for the AGN and comparison stars was 5\arcsec, and an annulus of $r=15\arcsec-20\arcsec$ was used to measure the sky background. For each filter, data points measured from multiple images taken at an individual telescope during a night were averaged to produce a single data point in the final light curve, but data points from different telescopes were not averaged together. All light curves in a given filter from different telescopes were intercalibrated using the code PyCALI \citep{Li2014}, which also expands the measurement uncertainties on data points to account for the systematic intercalibration uncertainty for each telescope's data. The data were flux calibrated using comparison star magnitudes from the APASS catalog \citep{APASS_DR10}. The total number of data points in the final combined light curves is  260 ($g$), 296 ($r$), 278 ($i$), and 240 ($z$).  The  optical light curve data are listed in Table \ref{table:photdata} and displayed in Figure \ref{fig:contlc}.

\section{Results}
\label{sec:results}

\subsection{Continuum lag results}
\label{sec:lags}

\begin{figure}
    \centering
    \includegraphics[width=\columnwidth]{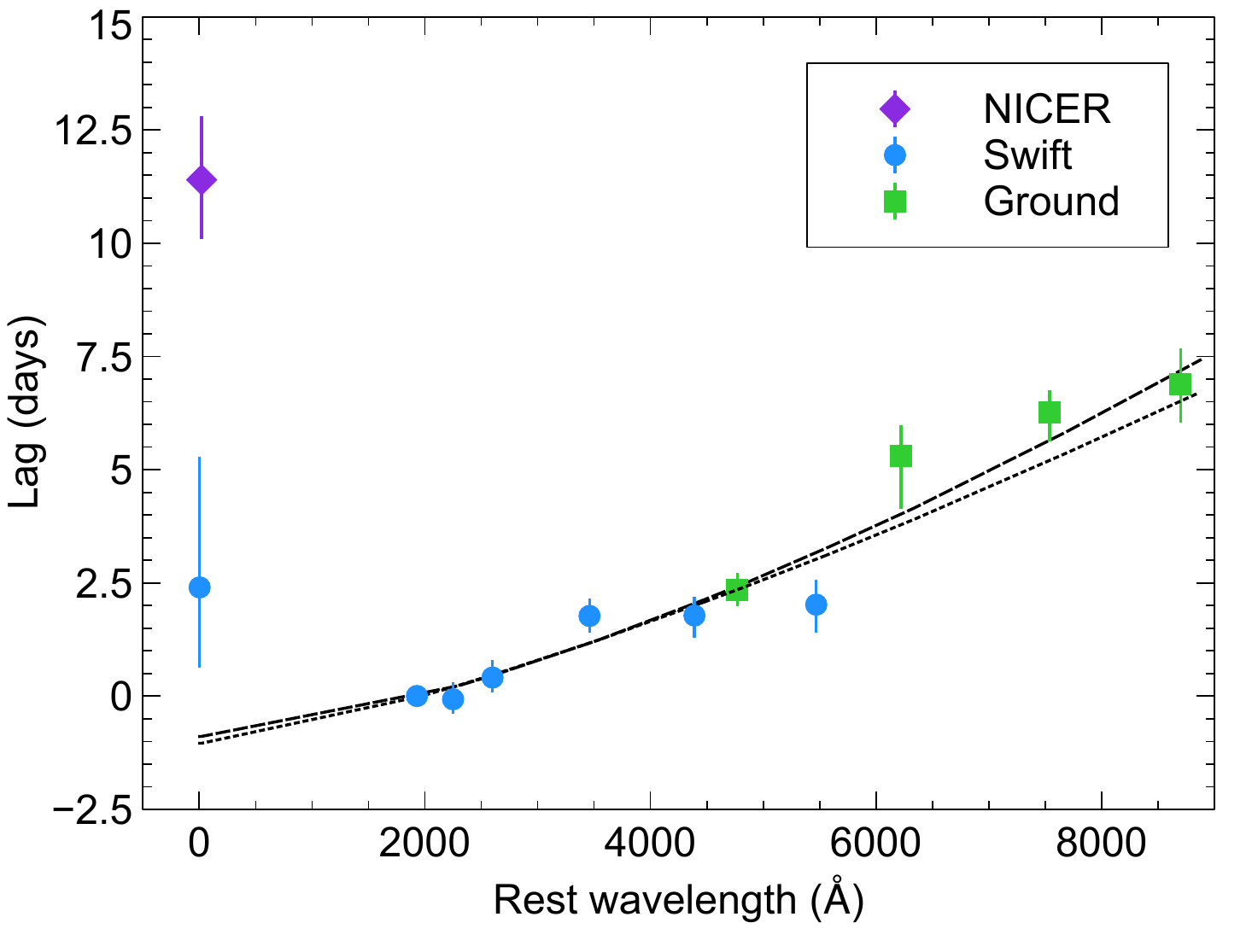}
    \caption{Observed continuum lags calculated with respect to the Swift/UVW2 band. The NICER soft X-ray lag is shown as a purple diamond; Swift as blue circles and ground-based $griz$ bands as green squares. The dotted and dashed lines show the best-fitting $\tau\propto\lambda^{\beta}$ relations, where $\beta$ is fixed at the canonical value of 4/3 for the dotted line, and allowed to be free for the dashed line ($\beta=1.5\pm0.4$). While the canonical $\tau\propto\lambda^{4/3}$ relation provides a good fit, the normalization of the lag is almost 10x larger than expected given the mass and accretion rate of this object. The Swift soft X-ray point has been removed for clarity, but is consistent with the NICER lag, just with a much larger error bar; see Table~\ref{table:contlags} for exact numbers. }
    \label{fig:contlags}
\end{figure}

\begin{deluxetable}{lccC}
\label{table:contlags}
\tablewidth{0pt}
\tablecaption{Continuum lags and light curve properties. Lags are computed with respect to the {\it UVW2} reference band, where positive indicates a lag behind {\it UVW2}.}
\tablehead{
\colhead{Filter/Band} &
\colhead{Telescope} &
\colhead{$R_{\rm max}$} &
\colhead{Lag centroid} \\
 & & & \colhead{(days)}}
\startdata
Hard X-ray (1.5--10~keV) & Swift &  0.68 & 2.4_{-1.8}^{+2.9}\\
Soft X-ray (0.3--1.5~keV) & Swift &  0.58 & 10.4_{-4.3}^{+3.0}\\
Soft X-ray (0.3--1.5~keV) & NICER &  0.62 & 11.4_{-1.3}^{+1.4}\\
{\it UVW2} (1928\AA) & Swift &  1.00 & 0.00\pm0.2\\
{\it UVM2} (2246\AA) & Swift &  0.95 & -0.07_{-0.32}^{+0.37} \\
{\it UVW1} (2600\AA) & Swift &  0.95  & 0.41_{-0.32}^{+0.38}\\
{\it U} (3465\AA) & Swift &  0.94  &  1.77_{-0.38}^{+0.39}\\
{\it B} (4392\AA) & Swift & 0.89  &  1.78_{-0.50}^{+0.42} \\
{\it g} (4770 \AA) & Ground &  0.95 & 2.34_{-0.36}^{+0.38}\\
{\it V} (5468\AA) & Swift &  0.83  & 2.02_{-0.62}^{+0.55}\\
{\it r} (6215\AA) & Ground & 0.87 & 5.31_{-1.17}^{+0.67}\\
{\it i}	(7545\AA) & Ground & 0.88 & 6.27_{-0.65}^{+0.49}\\
{\it z}	(8700\AA) & Ground & 0.76 & 6.88_{-0.83}^{+0.79}\\
\enddata
\end{deluxetable}


We calculate lags with PyCCF \citep{pyccf}, using the standard linear Interpolated Cross-Correlation Function (ICCF) technique, where uncertainties are estimated with the flux randomization, random subset sampling approach \citep[as implemented by][]{peterson04}. This entails creating 1000 realizations of the light curves, where each flux point is drawn randomly from a Gaussian distribution with mean and standard deviation equal to the measured flux and its 1$\sigma$ uncertainty. In addition, the light curve is randomly resampled with repetitions, i.e., some points are selected multiple times, while others are not selected at all.  For each realization we measure the cross-correlation function (CCF) and its centroid value.  The lag is the median of the CCF centroid distribution, and its 1$\sigma$ uncertainty from the 16\% and 84\% quantiles. The time lag range considered in the CCF was $\pm 30$~days. Because it probes closest to the thermal peak of the hot accretion disk, and for consistency with previous results, we use the Swift/$UVW2$ band as the reference light curve.

The observed light curves and resulting rest-frame lags (over the period HJD=2458770--2458885) are shown in Fig.~\ref{fig:contlc} and details of the lag centroid and maximum correlation coefficient $R_{\rm max}$ are shown in Table~\ref{table:contlags}.  The right-hand panels of Fig.~\ref{fig:contlc} show the CCFs (solid lines) and the CCF centroid distributions. Fig.~\ref{fig:contlags} shows the lags as a function of wavelength.  The UVOIR lags increase with wavelength, approximately following $\tau \propto \lambda^{4/3}$, as expected for a standard Shakura \& Sunyaev thin disk \citep{Cackett2007}.  We fit the UVOIR lags with the function: $\tau = \tau_0 \left[ \left(\lambda/\lambda_0\right)^\beta - y_0 \right]$, with $\lambda_0 = 1869$ \AA\ (the rest-frame wavelength of the UVW2 band), and $\beta = 4/3$, and where $y_0$ allows the fit to cross zero lag at $\lambda=\lambda_0$.  This gives a best-fitting value of $\tau_0 = 1.11\pm0.06$ days. Allowing $\beta$ to be free yields consistent results, with $\beta = 1.5\pm0.4$. If we perform the ICCF analysis on a subset of the campaign from HJD 2458770--2458850 (i.e., removing the flare at the end), we find a very similarly shaped lag-wavelength relation,  but the normalization is smaller: $\tau_0 = 0.62\pm0.08$ days. Regardless of whether we consider the flare data or not, the normalization of the lag-wavelength relation is quite large, and will be discussed more in Section~\ref{sec:discuss}. 

The reprocessing model provides an adequate fit to the data, with a few noteworthy exceptions, described in the following paragraphs. Nearly all intensive reverberation campaigns that utilized Swift and ground-based monitoring \citep{edelson19,cackett18,vincentelli21,Hernandez2020} have found significant excess lags in the $u$ band and $r$ band, relative to the adjacent bands or to the model fits. Mrk~335 is no exception. However, a systematic analysis of four AGN with intensive Swift monitoring campaigns showed typical $u$ band excesses of a factor of 2.2 on average \citep{edelson19}. Mrk~335, on the other hand, shows an excess of the $u$ and $r$ bands of only 30\% greater than expected. See Section~\ref{sec:discuss} for more details.

The other obvious outlier to the $\tau \propto \lambda^{4/3}$ relation in Fig.~\ref{fig:contlags} is the soft X-ray lag. In the standard reprocessing picture, the X-rays are the driving light curve and therefore should lead the other light curves, but here we see that the soft X-rays are delayed with respect to the UVW2 by $\sim 10$ days. This $\sim 10$-day delay is seen both with NICER and with Swift, though both show a low soft X-ray vs. UVW2 maximum correlation coefficient of $R_{\rm max} \sim 0.6$. The low Swift-UVW2 correlation coefficient is likely due to the fact that Swift did not observe the flare at high cadence. To understand where this soft X-ray lag originates, we compare the NICER soft X-ray light curve to the $g$-band light curve (which is the best sampled optical light curve, and importantly, covers the flare period). The correlation coefficient between NICER soft X-rays and the $g$-band light curve is much higher ($R_{\rm max} = 0.80$) with an X-ray lag of $3.5^{+1.1}_{-0.9}$ days.

 \begin{figure}
\centering
\subfigure{\includegraphics[width=\columnwidth]{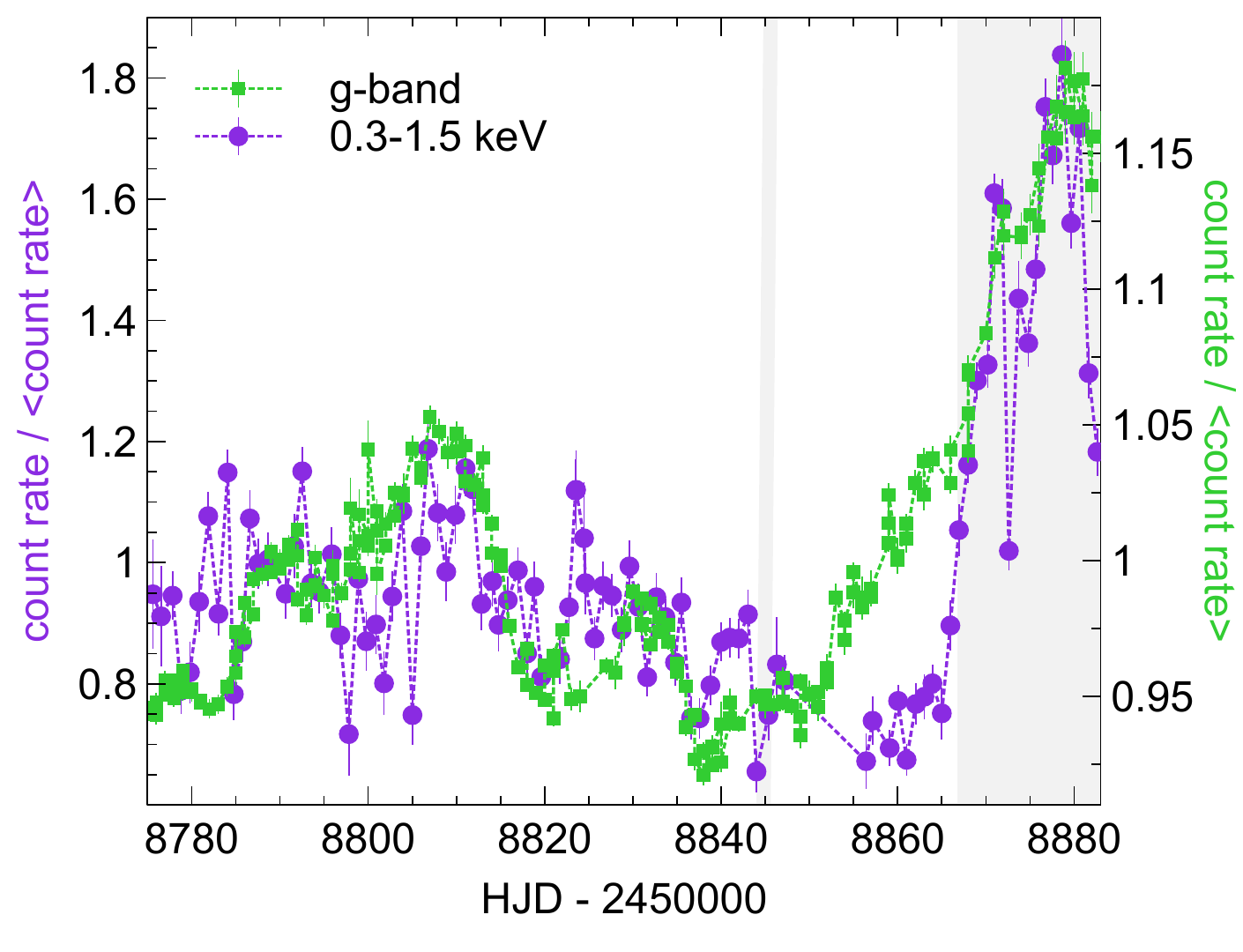}}
\caption{Comparison of the NICER 0.3--1.5~keV soft X-ray light curve to the $g$-band light curve, shown to highlight the late time rise of the X-rays relative to longer wavelengths. Notice that the X-ray and optical light curve variations are shown on different axes, with the X-ray varying more than the optical. The gray shaded regions indicate the times from which the spectra shown in Fig.~\ref{fig:lc-spec-spec} were taken.}
\label{fig:lc-spec-lc}
\end{figure}


We tested whether the 10-day soft X-ray lag is due to variability across the entire campaign or simply driven by the large flare at the end of the campaign seen both in X-rays and UVOIR. We performed the ICCF analysis on a subset of the campaign from HJD 2458770--2458850 (i.e., removing the flare at the end). While the NICER lags UVW2 slightly by $2.9\pm2$~days, $R_{\rm max}$ value is very low ($R_{\rm max}=0.47$), and so ought not be taken with too much credence.

We conclude that the soft X-ray lag is mostly an artifact of the large flare at the end of the campaign, where the X-rays rise after the UVOIR bands. This can be seen most directly by simply overplotting the NICER soft X-ray light curve and the $g$-band light curve (Fig.~\ref{fig:lc-spec-lc}). We use the $g$-band light curve simply because it is the best sampled light curve. Similar conclusions would be drawn from comparison with the UVW2 light curve. We compare the light curves over the same timescale (HJD 2458776-2458883), and overplot them by dividing each light curve by its mean. One can see that the correlation from the non-flare period is low (as discussed in the previous paragraph), but the flare shows a distinct energy dependence. The X-rays rise $\sim 15$ days after the rise of the $g$-band. This is counter to the canonical reprocessing scenario where the optical rises in response to a rise in the X-ray light curve. The standard reprocessing scenario can be salvaged if the X-ray luminosity we observe is not the same as the X-ray luminosity `seen' by the disk. We will discuss possible interpretations for the soft X-ray delay in Section~\ref{sec:discuss}.

\subsection{Spectral evolution}
\label{sec:spec}

 
 \begin{figure}
\centering
\subfigure{\includegraphics[width=\columnwidth]{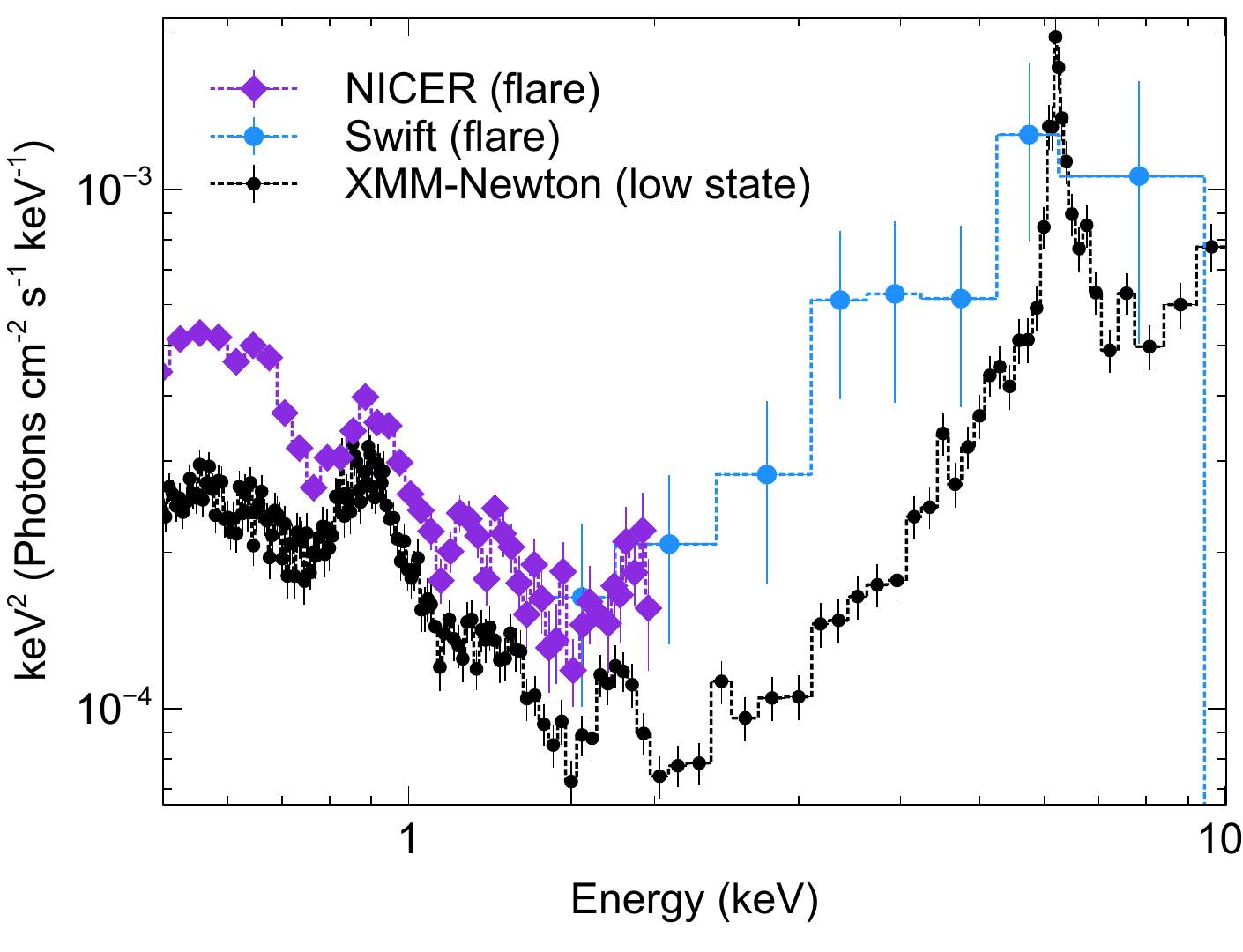}}
\caption{The XMM-Newton spectrum (black) was taken during one of the lowest points of the campaign (HJD~2458845), and the NICER (purple) and Swift (blue) spectra are from the flare state (integrated over HJD~2458867 to HJD~2458883). This highlights that the flare was not simply seen in soft X-rays alone, as would be expected from a decrease in the level of obscuration.}
\label{fig:lc-spec-spec}
\end{figure}


Next we compare the X-ray spectra from the low-flux state to the flare state. During the low-flux state of our 100-day campaign, XMM-Newton took one observation (denoted as thin gray line at HJD=2458845 in Fig.~\ref{fig:lc-spec-lc}). This spectrum is shown in black in Fig.~\ref{fig:lc-spec-spec}, and is very similar in shape and flux to low-state observations taken one year earlier (see \citealt{parker19} for a detailed spectral analysis of those observations). The low energies are dominated by narrow photoionized emission lines from circumnuclear material \citep{liu2021}, and a prominent neutral iron~K line is present at 6.4~keV, presumably from further out in the dusty torus. 

For comparison, we overplot the NICER and Swift spectra from the flare at the end of the campaign, from HJD~2458867 to HJD~2458883. The overall flux level has increased by a factor of $\sim2.3$, and most importantly, the relative change in hard X-rays is {\em larger} than that in soft X-rays, which cannot be explained by a change in obscurer column density alone. This larger relative increase in hard X-rays can also be seen by comparing the Swift soft and hard band light curves in Fig.~\ref{fig:contlc}. The origin of the X-ray flare is discussed in Section~\ref{sec:discuss}.

\begin{figure}
    \centering
    \includegraphics[width=\columnwidth]{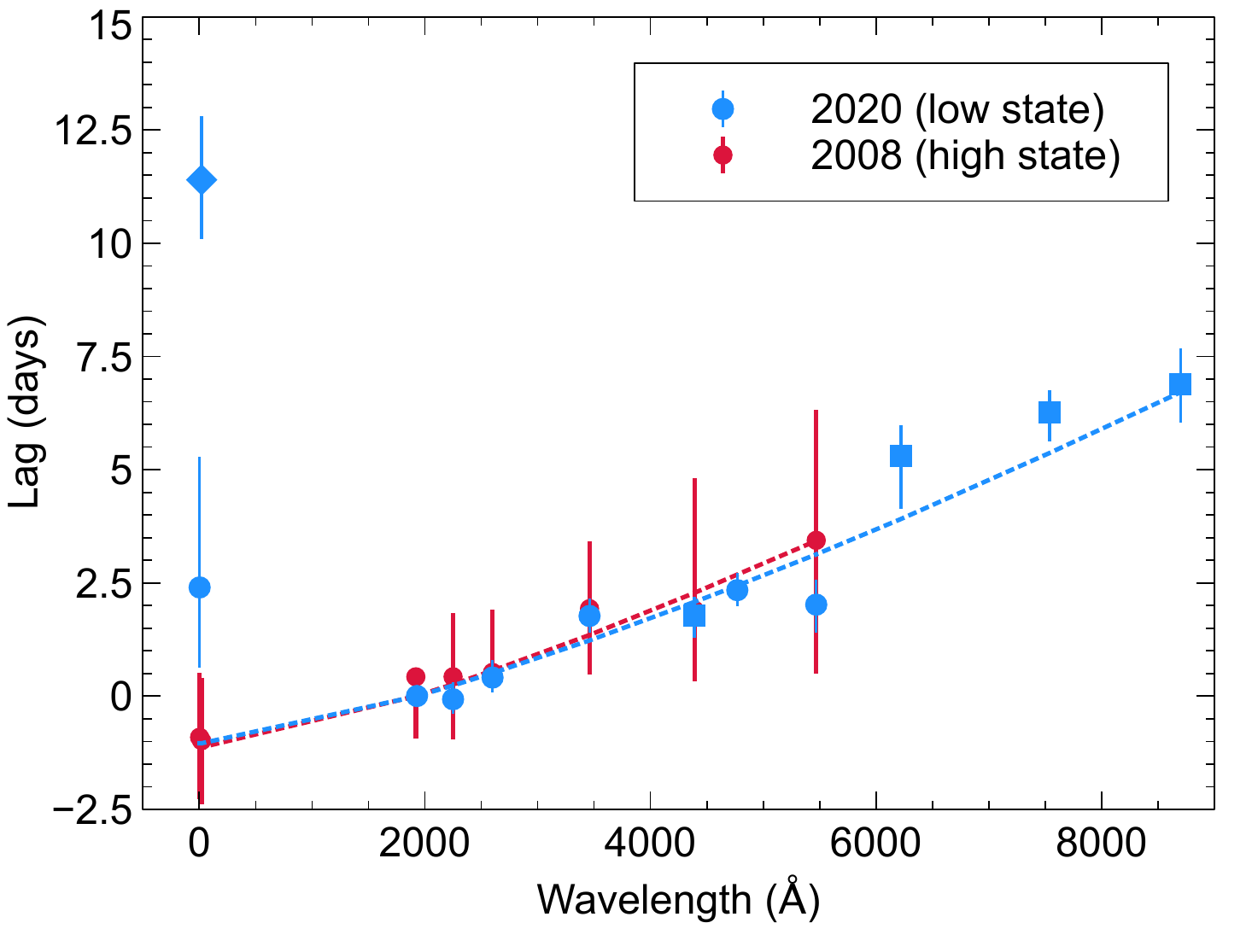}
    \caption{Lag vs. wavelength, comparing our recent intensive reverberation mapping campaign in the X-ray low state (blue points) to a high-state campaign in 2008 (red points). The marker shape indicates the facility (diamond for NICER, circle for Swift and square for ground based). The error bars from the high state are much larger because of the lower cadence observations, but despite this, one can see that the UVOT lags behave the same in the high X-ray state compared to the low-state. There is no evidence for an X-ray lag in the high state.}
    \label{fig:archival_lag}
\end{figure}

\subsection{Continuum lags in an archival high-flux state}
\label{sec:lag_compare}

Mrk~335 is one of the best studied AGN in the sky, which allows us to compare the results of our campaign to previous observations. As highlighted in Fig.~\ref{fig:swiftlc}, Swift has monitored Mrk~335 in X-rays and UVW2 at a roughly weekly cadence since 2008. From 2008--2009, there was a particularly high-cadence campaign of monitoring every 1--3~days in X-rays and all UVOT filters (\citealt{grupe12}, and see red shaded region in Fig.~\ref{fig:swiftlc}). This campaign took place at a particularly high-flux state, when the X-rays were $>10$ times brighter and the UVW2 $\sim 1.5$ times brighter than during our campaign. While not as high-cadence as our recent campaign, it does provide an interesting comparison to our low-flux campaign. 

We perform the same ICCF analysis described above on the archival dataset. Similar to \citet{gallo18}, we find marginal evidence that the X-rays lead the UVOIR bands. Fig.~\ref{fig:archival_lag} shows a comparison of the lag vs. wavelength for the high-cadence campaign (low-flux) and archival campaigns (high-flux). While the error bars are much larger in the high-flux state, the UVOIR lags do appear consistent with the low-flux campaign. The $\sim 10$~day soft lag is not present in the archival high-state data, and in fact, both soft and hard X-rays are consistent with expectations from the simple reprocessing picture where lags scale as $\tau \propto \lambda^{4/3}$ (dashed line, Fig.~\ref{fig:archival_lag}). \citet{komossa20} also report a 1.5~day UV/optical lag (consistent with these results) after the X-rays began to rise again in late 2020. These results suggest that the UV/optical lags do not depend strongly on mass accretion rate.

\section{Discussion}
\label{sec:discuss}

The  inflow of gas through a geometrically thin accretion disk is driven by local disk instabilities that result in variability of the optical and then UV emission as the fluctuations propagate inwards on a viscous timescale (tens to thousands of years at $\sim 1000~\Rg{}$). Our months-to-year monitoring campaigns commonly observe inter-day variability where the UV {\em leads} the optical by a few days or less. This rapid variability and short lags can  be explained in a reprocessing scenario where emission from the central regions irradiates and heats up the accretion disk at larger radii. The X-rays are the natural choice for the driving light curve as X-rays are known to originate in the central regions and vary rapidly. However, in several sources, the observed X-ray/UV correlation is lower than predicted ($R_{\rm max}<0.75$), and much weaker than the cross-correlations between UVOIR bands ($R_{\rm max}>0.8$). 

It has been suggested that the low X-ray/UV correlation means that the X-rays are not the driving light curve at all \citep{Gardner2017,Mahmoud20}. Moreover, in some cases, it is clear that absorption along our line of sight due to obscuring disk winds or the AGN torus can complicate our inference of the intrinsic continuum and its echoes, as in the recent AGN STORM 2 campaign of Mrk~817 \citep{kara21}. The origin of the driving light curve (and its implications for the inferred geometry and energetics of the corona and inner accretion flow) remains one of the biggest outstanding problems in accretion disk reverberation mapping. 

Despite the campaign taking place in a low X-ray flux state, Mrk~335 shows a relatively high correlation between X-ray and UV (especially between hard X-rays and UV). Moreover, the UVOIR lags appear to show the same signatures of disk reverberation in both high- and low-X-ray flux states. Similar to Mrk~817, in its X-ray low-flux state, Mrk~335 did display broad blueshifted UV absorption troughs from an outflow at the inner BLR \citep{Longinotti2019, parker19} that is likely also responsible for some obscuration in X-rays \citep{parker19}. But our 100-day campaign still reveals rapid X-ray variability (even in hard X-rays), which indicates that there is still some intrinsic variability from the corona. Therefore, while obscuration is likely important in understanding the long-term evolution of the X-ray light curve (e.g. see \citealt{komossa20}), we still see rapid variability from the central source that can explain the UVOIR rapid variability and subsequent reverberation lags. 

\begin{figure}
    \centering
    \includegraphics[width=\columnwidth]{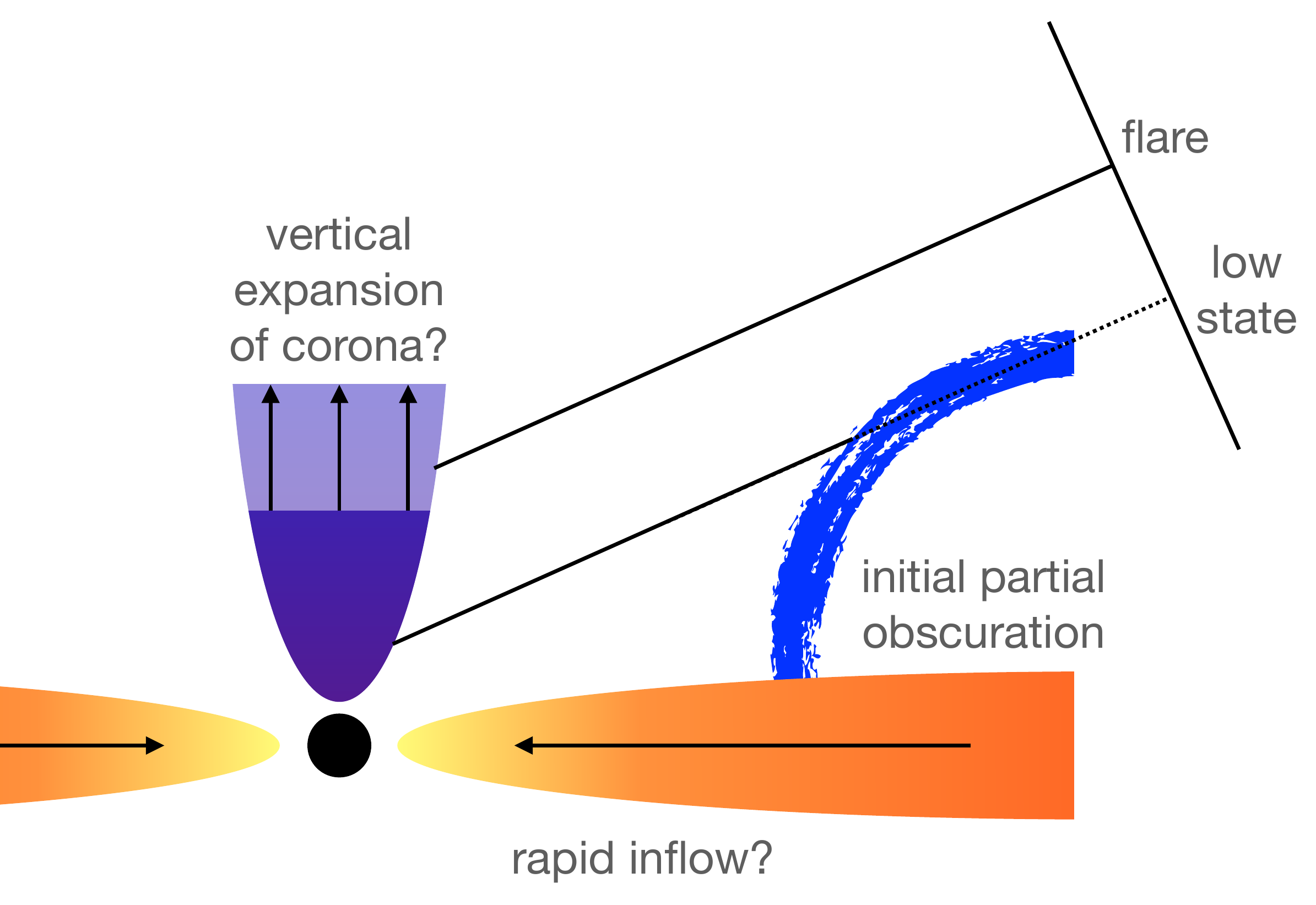}
    \caption{Schematic demonstrating potential scenarios to explain (1) the broadband flare at the end of the campaign (Section~\ref{sec:flare_disc}), and (2) the unusually long lags (Section~\ref{sec:long_lag_disc}). HST observations have indicated the presence of an obscurer in the low-flux state \citep{Longinotti2019,parker19}. Our high-cadence campaign took place at the end of this state, and potentially caught the initial rise out of the low-flux state. The broadband flare at the end of the campaign starts in UV/optical, followed by a rapid rise in X-rays. This could be due to the inflow of mass accretion rate fluctuations that trigger rapid X-ray flaring, or in a reprocessing scenario, could result in the late-time expansion of the corona, so the disk sees the rise in X-rays before the observer does. Such an extended corona scenario might also explain the anomalously long UVOIR lags.}
    \label{fig:schematic}
\end{figure}

In the following sub-sections, we discuss two observational puzzles arising from this campaign that may provide important clues on the nature of the X-ray corona: (1) How do we explain the late-time  multi-wavelength flare at the end of the campaign, and (2) Why are the observed disk lags exceptionally long in this source? See Fig.~\ref{fig:schematic} for a schematic to aid the discussion.

\subsection{On the origin of the broadband flare}
\label{sec:flare_disc}

High amplitude flares are to be expected in a stochastic noise process, and so while a flare itself is perhaps not unusual, it does provide us with a unique opportunity to probe the impulse-response function of the disk in a new way (i.e. without a cross-correlation analysis). In our case, the flare begins in the optical/UV bands, but no increase is seen in either NICER or Swift. Then in the final 2 weeks of the campaign (unfortunately, just as the NICER background also increased due to optical loading), the NICER and Swift light curves indicate that the X-rays also rise (with a soft X-ray spectrum similar to that seen in low-state XMM-Newton spectra, and therefore not likely to be associated with the background). This apparent X-ray rise occurs $\sim 15$~days after the UV rise (Fig.~\ref{fig:lc-spec-spec}). Swift indicates that the flare in X-rays is stronger in hard X-rays than in soft, suggesting that the flare is intrinsic to the corona, and not due to a lower covering fraction or column density of an obscurer (Fig.~\ref{fig:lc-spec-spec}). Moreover, after our 100-day low-state campaign (once the source became visible again $\sim 6$ months later), the X-rays had clearly risen from their deep low-state \citep{komossa20}. Here, we explore potential scenarios to explain the re-emergence of the X-rays after the low-flux state, and the initial rise in UV/optical.

The initial rise in UVOIR is a much gentler and lower amplitude than that in X-rays, which means that formally, there can be no linear response function that can be convolved with the UVOIR light curve to produce the delayed X-ray light curve. This may corroborate the idea that the stochastic nature of AGN light curves could lead to coincidental flares in multiple bands, or could indicate a non-linear response is at play. Such effects could arise, for instance, from the inflow of mass accretion rate fluctuations that initially start far out in the accretion disk, propagate inwards, and eventually build up to the innermost regions. This may trigger rapid magnetic reconnection events close to the black hole that result in a sudden flash of X-rays. 

Such scenarios have been proposed to explain the long timescale ($\sim$year) X-ray/optical correlation, which deviates from expectations from reverberation observed on short timescales on days--weeks. For instance, \citet{uttley03} found that the X-ray variability amplitude is smaller than the optical on long timescales, and \citet{arevalo08} found the opposite was true on short timescales. This could be explained if the optical drives the variability on long timescales (through propagation affects), but reverberation dominates on short timescales (where X-rays are the driving light curve). Most recently, 
\citet{Hernandez2020} showed that the long timescale variability shows the optical leading the UV (again, as expected from propagation lags), but on short timescales, the lags scale as $\tau \propto \lambda^{4/3}$, as expected for reverberation. A similar model is described in the recent analysis of \citet{neustadt22}.

In our case, if we are observing a single mass accretion rate wave at the end of the campaign, the timescale of $\sim 15$~days is very fast for a viscous timescale. In fact, it is closer to the dynamical timescale, so perhaps this flare is a pressure wave or magnetic wave moving at the timescale for the disk to vibrate in response to some mechanical disturbance.

Alternatively, the late response of the X-rays could also be explained if the luminosity of the corona increases with increasing spatial extent, as suggested by some X-ray reverberation studies (e.g., \citealt{Alston2020}). Even in Mrk~335, X-ray spectral modeling of the emissivity profile of the iron~K emission line \citep{wilkins15} and other reflection features \citep{gallo19} suggest that some flares may be due to a vertical expansion of the corona. If, at early times, the intrinsic X-ray luminosity is low and the corona is more spatially compact, an equatorial wind could obscure the corona. Then, in the final days of the campaign, as the corona luminosity increases and it becomes more spatially extended, this will increase the amount of irradiation of the disk (causing the upturn in UV emission), and finally we have a less obscured sight-line to the central source. Such a scenario would predict that the hard X-ray light curve (1.5--10~keV) would rise before the soft, since the hard X-rays are less impacted by line-of-sight obscuration, but unfortunately, this cannot be tested with the current campaign as NICER did not detect hard X-rays above the background. This interpretation of an expanding corona during the flare may also explain the large amplitude of the lags, as discussed in the next section.

\subsection{On the very long X-ray/UVOIR lags}
\label{sec:long_lag_disc}

In Section~\ref{sec:lags} we found that the shape of the UVOIR lag-wavelength relation roughly follows the $\tau \propto \lambda^{4/3}$ relation expected for reverberation lags echoing off a standard thin disk. Fitting this function to the data, we find a normalization of $\tau_0 = 1.11\pm0.06$ days in the low X-ray flux state, and nearly identically shaped UVOIR lags in the high-flux state. Excluding the flare from the analysis results in a lower normalization of $\tau_0 = 0.62\pm0.08$ days.

Following Equation~12 of \citet{fausnaugh16}, we compare the $\tau_0$  normalization of the lag-wavelength relation to expectation for a standard Shakura \& Sunyaev thin disk, where the disk temperature profile scales as $T \propto R^{-3/4}$ \citep{Cackett2007}. As in \citet{fausnaugh16} and other broadband reverberation campaigns (e.g., \citealt{cackett20}), we assume a $T$ to $\lambda$ conversion factor of $X=2.49$, radiative efficiency of $\eta=0.1$, and $\kappa=1$ for equal heating of the disk by X-rays and viscous affects. With these assumptions in place, the normalization of the lags is a function of black hole mass and the bolometric luminosity. We assume a black hole mass of $(2.6 \pm 0.8) \times 10^{7} M_{\odot}$, based on previous broad line region reverberation mapping campaigns of Mrk~335 \citep{grier12}\footnote{An important caveat here is that the optical reverberation lags, from which this mass was derived, were measured with respect to the 5100~\AA\ continuum, assuming it all originates in the accretion disk. If there there is a contribution from the diffuse continuum emission from the BLR \cite{koristagoad01} within this band, then the inferred mass would be larger.}. For $L/L_{\mathrm{Edd}}=0.07$ \citep{tripathi20}, we anticipate a lag-wavelength normalization of $\tau_0 = 0.1$ days, 11 times shorter than our measured $\tau_0$ of 1.1~days (for the entire campaign, including flare). Excluding the flare at the end of the campaign, leads to a lower $\tau_0$ of 0.6~days, bringing the discrepancy between observation and prediction to a factor of 6. Even if we make the rather extreme and unlikely assumption that Mrk~335 is actually Eddington-limited ($L/L_{\mathrm{Edd}}=1$) despite its low-flux state, we still predict a lag normalization that is $\sim 2.4-4.3$ times smaller than what is observed. It is not unusual to find UV/optical lags that are `too long' (e.g. \citealt{fausnaugh16, cackett18, edelson19}), but usually the measured lags are $\sim 3$ times longer than expected. This is the largest discrepancy with standard disk theory to date. Perhaps the fact that removing the broadband flare from the analysis puts our observed lags more in line with previous AGN suggests that the flare is dominated by processes other than disk reverberation.

One proposed solution to the `too-long-lag' problem is that the measured lags are not simply due to light travel times from the corona to the accretion disk, but also there is a contribution from diffuse continuum emission (DCE) from the BLR, due to free-free and free-bound hydrogen transitions \citep{koristagoad01,koristagoad19}. The light travel time from the inner regions out to the BLR is much longer, and therefore even a small contribution of DCE in the UV/optical emission will produce measured lags that are longer than expected from just a pure accretion disk. In addition to DCE contributing to all UV/optical continuum bands, \citet{koristagoad01} predict an excess lag at the $u$ (which contains the Balmer jump), and the $r$ band (which contains the H$\alpha$ line). Indeed, \citet{cackett18}, using cadenced spectroscopic observations, showed that in NGC~4593 this $u$ band excess was, in fact, a broad excess leading up to the Balmer jump. Moreover, recent timescale-resolved analysis showed that by removing the long-timescale variability from the lag analysis, the $u$ band excess disappeared, supportive of the idea that the $u$ band has a large contribution from the DCE much further (and therefore more slowly varying) than the disk reverberation lags \citep{cackett22}.

Here, in Mrk~335, if the DCE is to remedy this extreme `too-long-lag' problem, then this would imply that there should be a larger than usual contribution from the BLR in the UV/optical lags, and therefore, also a larger excess in bands where the DCE contributes. In a systematic analysis of four AGN with intensive Swift monitoring campaigns, \citet{edelson19} found that the $u$ band exceeds the lag expected by the $\tau \propto \lambda^{4/3}$ relation by a factor of 2.2 on average. In Mrk~335, we also see an excess of the $u$ and $r$ bands, but both bands are only 30\% greater than expected. At face-value this appears to present a challenge for the DCE model, but it is important to recognize that at the redshift of Mrk~335, in addition to the $u$ band containing the Balmer jump, several other bands also contain contributions from the BLR, namely: the $g$ band is centered on He~II and also contains $H\beta$, $r$ band contains He~I 5876 and $H\alpha$ and $z$ contains the Paschen jump. Therefore, because the DCE component may contribute to the lags in several bands, it could reduce the contrast between the U band excess lags and those lags at longer wavelengths. Detailed photoionization modeling and timescale-resolved analysis will help in isolating the contributions from the disk, the DCE and other variability processes.

Alternatively, the long lags could be attributed to the primary irradiating source being further from the disk than initially expected. If the corona is extended out to $\sim 100$\Rg{}, the X-ray source will preferentially irradiate larger radii in the disk, making the response functions wider, and thus the UV/optical time lags longer (see Eq.~1 of \citealt{kammoun21b} for more details). Recently, these authors modeled the X-ray/UVOIR lags of several sources and \citet{panagiotou22} modeled both the time lags and power spectra in NGC~5548 with general relativistic ray tracing simulations \citep{dovciak22}, and constrained the height of the corona to be somewhere between $\sim 7-70$~\Rg{} for previously studied AGN. This accounts for the factor of 2--3 longer lags than expected (where the irradiating source is assumed to be coplanar with the disk). Mrk~335 shows lags that are 6--11 times the expected amplitude, and so may indicate a corona that is even more extended than other AGN in the sample. Interestingly, of all the AGN with intensive X-ray/UVOIR monitoring campaigns thus far, Mrk~335 has the clearest indication of a relativistically broadened iron line in the X-ray band (e.g. \citealt{parker14,wilkins15}) and short X-ray reverberation lags \citep{kara13}, both of which require a corona that is close to the black hole ($\sim 10$\Rg{}). Therefore, to reconcile such observations, the corona needs to be vertically extended, with a base close to the black hole. 

These long X-ray/UVOIR lags may be further evidence of a vertically extended corona in Mrk~335, as suggested by earlier X-ray spectral studies \citep{wilkins15, gallo19}. In this picture, the fact that the $\tau_0$ normalization is smaller if we exclude the flare from the analysis, would suggest that the corona is more compact in the early part of the campaign (when the X-ray flux is lowest), and increases in spatial extent during the flare. The vertically extended corona may represent the base of the bi-polar parsec-scale jet recently spatially resolved with Very Long Baseline Array observations of Mrk~335 \citep{yao21}.


\section{Conclusions}
\label{sec:conclusions}

To summarize, our major findings of the 100-day Swift/NICER/ground based reverberation mapping campaign of Mrk~335 are: 
\begin{itemize}
     \item Despite the campaign occurring in an unprecedented low X-ray state, the UVOIR lags appear unaffected, and the wavelength dependence of the lags is consistent with expectations of reverberation delays off an accretion disk (Fig.~\ref{fig:contlags}). 
    \item Archival observations taken in a high-flux state show very similar UVOIR lags as those found in the low X-ray state (Fig.~\ref{fig:archival_lag}).
    \item Given its black hole mass and accretion rate, the amplitudes of the UVOIR lags (in both low and high state) are 6--11 times larger than expected from a standard Shakura-Sunyaev disk \citep{fausnaugh16}. 
    \item The $u$ and $r$ band excess lags (commonly attributed to contamination from the BLR; \citealt{koristagoad01}) are present, but are less prominent than typically seen in other AGN (e.g. \citealt{edelson19}; Fig.~\ref{fig:contlags}).
    \item The X-rays show little variability, and do not correlate highly with the UVOIR bands. At the end of the campaign, the UVOIR bands reach their highest level, and though lower significance (due to limited singal-to-noise and lower cadence), the X-rays also appear to rise  (Fig.~\ref{fig:contlc}). This X-ray rise occurs after the UVOIR rise, contrary to the standard reprocessing picture (Fig.~\ref{fig:lc-spec-lc}).
    \item The overall normalization of the X-ray spectrum rises at the end of the campaign (Fig.~\ref{fig:lc-spec-spec}), which suggests that the flare is not simply due to lower absorption effects, but rather, is an intrinsic increase in the luminosity of the corona.
    \item AGN are known to vary as a stochastic red-noise variability process, but if we interpret the soft X-ray delay physically, it could indicate either mass accretion rate fluctuations propagating inwards in the flow on timescales faster than the local viscous time, and/or that the spatial extent of the corona increases at the end of the campaign. The latter interpretation of an extended corona may also explain the unusually long UVOIR lags (Fig.~\ref{fig:schematic}). 
\end{itemize}

\section*{Acknowledgements}

 EK thanks the \xmm\ Science Operations Center Coordinators, Jan-Uwe Ness and Lucia Ballo, and \nustar\ Science Operations Manager, Karl Forster, for their patience in organizing the XMM-Newton and NuSTAR observations, as we waited for the X-ray flux to increase.  EK thanks Daniel Proga for interesting discussions on the X-ray/UV connection, and Rick Edelson for assistance in organizing the campaign. This work made use of data supplied by the UK Swift Science Data Centre at the University of Leicester. 

This work makes use of observations from the Las Cumbres Observatory global telescope network. 
The Liverpool Telescope is operated on the island of La Palma by Liverpool John Moores University in the Spanish Observatorio del Roque de los Muchachos of the Instituto de Astrofisica de Canarias with financial support from the UK Science and Technology Facilities Council.

Some of the data used in this paper were acquired with the RATIR instrument, funded by the University of California and NASA Goddard Space Flight Center, and the 1.5-meter Harold L.\ Johnson telescope at the Observatorio Astron\'omico Nacional on the Sierra de San Pedro M\'artir, operated and maintained by the Observatorio Astron\'omico Nacional and the Instituto de Astronom{\'\i}a of the Universidad Nacional Aut\'onoma de M\'exico. Operations are partially funded by the Universidad Nacional Aut\'onoma de M\'exico (DGAPA/PAPIIT IG100414, IT102715, AG100317, IN109418, IG100820, and IN105921). We acknowledge the contribution of Leonid Georgiev and Neil Gehrels to the development of RATIR. 

EK acknowledges support from NASA grants 80NSSC20K0470 and 80NSSC20K0372, and is supported by the Sagol Weizmann-MIT Bridge Program. Research at UC Irvine was supported by NSF grant AST-1907290. EMC gratefully acknowledges support from the NSF through grant AST-1909199. YRL acknowledges financial support from the NSFC through grant Nos. 11922304 and 12273041 and from the Youth Innovation Promotion Association CAS. PD acknowledges support from NSFC  grant 12022301 and 11991051.
CH  acknowledges support from from NSFC grants  12122305 and 11991054. CSR thanks the UK Science and Technology Facilities Council (STFC) for support under the Consolidated Grant ST/S000623/1, as well as the European Research Council (ERC) for support under the European Union’s Horizon 2020 research and innovation programme (grant 834203). TL acknowledges support from the Zuckerman Postdoctoral Scholarship Program and by an appointment to the NASA Postdoctoral Program at NASA Goddard Space Flight Center, administered by Oak Ridge Associated Universities under contract with NASA.

\facilities{XMM-Newton, NICER, Swift, LCOGT, Liverpool:2m, Wise Observatory, Zowada, OANSPM:HJT}


\bibliographystyle{apj}
\bibliography{ref.bib}

\end{document}

%% file: defns.tex
\def\e0{\epsilon_0}

\def\ciii{\ifmmode {\rm C}\,{\sc iii} \else C\,{\sc iii}\fi}
\def\civ{\ifmmode {\rm C}\,{\sc iv} \else C\,{\sc iv}\fi}

\def\o5007{[O\,{\sc iii}]\,$\lambda5007$}

\def\nustar{{NuSTAR}}